# Probing dynamics in quantum materials with femtosecond x-rays

M. Buzzi, M. Först, R. Mankowsky, A. Cavalleri

*Max Planck Institute for the Structure and Dynamics of Matter, Hamburg, Germany*

*Optical pulses are routinely used to drive dynamical changes in the properties of solids. In quantum materials, many new phenomena have been discovered, including ultrafast transitions between electronic phases, switching of ferroic orders and non-equilibrium emergent behaviors such as photo-induced superconductivity. Understanding the underlying non-equilibrium physics requires detailed measurements of multiple microscopic degrees of freedom at ultrafast time resolution. Femtosecond x-rays are key to this endeavor, as they can access the dynamics of structural, electronic and magnetic degrees of freedom. Here, we cover a series of representative experimental studies in which ultrashort x-ray pulses from free electron lasers have been used, opening up new horizons for materials research.*

The equilibrium functional properties of solids are determined by the interplay between many microscopic degrees of freedom. These include the crystallographic structure, the arrangement of charges, spins, and orbitals as well as their dynamical fluctuations. The strong interactions between these many degrees of freedom create complex energy surfaces and make the ground state highly dependent on subtle differences in microscopic parameters, and on fine-tuning of external conditions. Understanding the origin of these "emergent" phenomena is, even at equilibrium, a formidable task that requires monitoring several degrees of freedom in the material at once. In the past two decades, equilibrium x-ray[1,2] and photoemission techniques[3,4] have provided an enormous amount of information and have contributed to understanding of equilibrium emergent states.

The present paper covers a new experimental direction in the physics of complex correlated electron systems, in which electromagnetic fields are used to control emergent properties away from thermodynamic equilibrium. Indeed, ultrashort laser pulses have been shown to be especially effective tools to manipulate magnetism[5,6] or ferroelectricity[7,8], to induce phase transitions at an ultrafast speed[9,10] and to trigger new emergent phenomena[10-13]. The underlying physics proceeds on femtosecond and picosecond timescales. Although these timescales have long been accessible with optical laser pulses since the 1970's, changes in the optical constants at visible and near infrared frequencies provide very limited information, only indirectly related to the microscopic degrees of freedom of interest.

Hence, complementary techniques that directly interrogate charge, spin and lattice degrees of freedom in the material are often applied to gain a deeper insight into the underlying physics. Time-resolved x-ray and electron diffraction, for example, directly track the photo-induced evolution of the crystal lattice (see Box 1), with some limitations in time resolution for electron diffraction experiments.[14] Time- and angle-resolved photoemission (tr-ARPES), on the other hand, is capable of tracking changes in the electronic band structure at different positions in the Brillouin zone. Similarly, knowledge of the transient element-specific local electronic structure can be gained from spectroscopic x-ray techniques, such as time-resolved x-ray absorption. Furthermore, tunable and intense x-rays provide the possibility of combining these spectroscopic techniques with the nano-scale spatial resolution given by diffraction and allow the direct study of the time-evolution of complex orders of charges, spins, and orbitals (see Box 1).

Ultimately, a comprehensive view of the physics at hand is obtained when combining results from different time-resolved techniques. In the following, we discuss examples where the use of x-ray techniques furthered our understanding of light-induced states of matter.

The development of ultrafast x-ray probes dates back at least two decades, made possible by the development of high intensity[15,16] amplified optical pulses[17,18]. Already in the 1990's, x-ray fluorescence from plasmas[19,20] spurred activity in this area. Despite the low flux and limited tunability of these first femtosecond x-ray sources, many rudimentary structural dynamics experiments were already reported at that stage, including studies of laser-induced disordering of organic films[21], photo-induced melting of semiconductors[22-24], detection of coherent acoustic[25,26] and optical[27] phonons, and photo-induced solid-solid phase transitions[28].

Other techniques combined the same high peak power femtosecond lasers with relativistic electron beams, first by exploiting 90-degrees Thomson scattering[29-31] and later by using lasers as energy modulators in electron storage rings[32-35]. These sources were tunable, and opened up femtosecond x-ray spectroscopies like ultrafast near edge x-ray absorption spectroscopy[36] or x-ray magnetic circular dichroism[37]. Accelerator based sources brought to the fore multiple-order-of-magnitude improvements in the x-ray flux[38,39], and culminated in the demonstration of x-ray from free electron laser (FEL) operation[40].

The present review covers the evolution of ultrafast materials research after x-ray FELs became available, focusing on how ultrafast x-ray diffraction and spectroscopy were used. We present a few representative experiments, especially those on ultrafast rearrangement of ferroic orders, of coupled charge, spin and orbital dynamics in complex oxides and in other strongly correlated materials. The notable case of photo-induced superconductivity will also be discussed, especially with respect to the contributions made by x-ray FEL experiments.

*Ultrafast Switching in Ferroelectric Materials*

Ferroelectric materials are of great scientific and technological interest, as they exhibit bi-stable, structurally distorted states of oppositely phased electrical polarization. For this reason, digital information can be stored in in ferroelectrics, making them interesting candidates for non-volatile memories. Typically, switching of the ferroelectric polarization is achieved by the application of pulsed electric fields. However, this reversal is based on incoherent dynamics and on the propagation of domain boundaries, which limits switching times to hundreds of picoseconds[41-43]. Several attempts to achieve ultrafast ferroelectric switching have been made, by driving the ferroelectric soft mode coherently with light pulses, either with impulsive Raman scattering[34,44-46] or direct excitation[8,47].

The properties of ferroelectrics can for example be controlled by the photoexcitation of charge carriers across the bandgap or by excitation of impurity levels. Such schemes have been used to facilitate polarization switching, control the domain nucleation and induce self-organized domain patterns[48-51]. The light-induced structural dynamics of ferroelectric $PbTiO_3$ thin films were shown to involve a distortion of the unit cell along the *c*-axis[52]. The response of the lattice constant along this direction was extracted from changes in the time-resolved x-ray scattering angle of an out-of-plane diffraction peak. A fast contraction of the film within the first 5 ps preceded a long-lived expansion, which was explained by dynamical charge screening of the depolarization field that in thin film ferroelectrics acts against the ferroelectric polarization[53].

More directly, the structural response of ferroelectric $Sn_2P_2S_6$ to the direct excitation of its soft mode was also measured with x-ray probes.[54] Following excitation with THz pulses of 120-kV/cm electric fields, coherent oscillations of the atoms along the soft mode coordinate were measured, through the corresponding modulation of the intensity of selected Bragg peaks. The amplitude of these motions corresponded *only* to a change of the ferroelectric polarization by 8%. The authors extrapolated that switching may become possible if the THz electric field was increased to 1 MV/cm.

In this context the development of mode-selective lattice control, so called "nonlinear phononics", has opened up new opportunities in ferroelectrics[55]. Transient reversal of the ferroelectric polarization[7] was observed using nonlinear optical probes, an important achievement that will likely motivate structural studies using ultrafast x-ray scattering of the involved atomic motions.

*Ultrafast magnetism in metallic systems*

It has long been known that ferromagnetism can be destabilized by light, although the use of optical pumping was shown to be able to drive the same phenomenon along a highly non-equilibrium path, and hence far faster than expected[6,56-58]. Time-resolved x-ray absorption spectroscopy (XAS) and x-ray magnetic circular dichroism (XMCD) were used to provide new insight into this class of processes, rejuvenating the field. For example, *Stamm et al.* combined these techniques to reconstruct the dynamical response in a nickel thin film excited with short near infrared pulses[37]. Within ~100 fs after optical excitation, the ferromagnetic order was completely quenched, indicating ultrafast transfer of spin angular momentum to auxiliary degrees of freedom, presumably the crystal lattice.

Measurements with simultaneous spatial resolution studied the growth of the magnetic disordering, evidencing the role of photo-excited electron diffusion on the length-scale of only tens of nanometers[59]. *Gutt et al.* demonstrated the use of single soft x-ray pulses from a FEL to record diffraction patterns from nanoscale magnetic-domain structures[60]. This approach yielded magnetic correlations with nanometer precision and 30 fs time resolution. *Pfau et al.* made use of magnetic small angle x-ray scattering to analyze modifications of the magnetic domain structure in a Co/Pd multilayer sample induced by near-infrared laser pulses[61]. These measurements revealed that the quench of spin angular momentum is accompanied by a decrease of the magnetic spatial correlations within the first few hundred femtoseconds. From its very high speed, it was speculated that this change could not result from domain wall motion, but was rather caused by spin-dependent transport of photo-excited electrons between neighboring ferromagnetic domains.

Magnetic x-ray scattering experiments have also played a key role in determining the importance of nanoscale inhomogeneities on ultrafast magnetization switching in GdFeCo, a collinear ferrimagnet with a strong magneto-optical response[62]. Near-infrared laser pulses were used to excite electrons on a time scale shorter than that of exchange interaction (~100fs), reversing the total magnetization of this compound after each pulse[6,63,64]. *Graves et al.* employed small angle x-ray scattering to gain insight on the mechanisms possibly responsible for ultrafast switching[5]. Figure 1a shows the chemical inhomogeneities in GdFeCo on a 10-nm length scale that divide the material into Gd- and Fe-rich regions. The magnetic scattering signal $S_q$ at low scattering momenta (q < 0.2nm$^{-1}$) allows for retrieving the average sample magnetization of the Gd and Fe spin sublattices, which are oppositely aligned at equilibrium. Figure 1b (top panel) showed that both net magnetizations are quenched rapidly within 1 ps after excitation. The magnetic scattering signal $S_q$ at high scattering momenta (q > 0.2nm$^{-1}$) is instead sensitive to the nanoscale structure of GdFeCo and showed significantly different dynamics (see Fig. 1b, bottom panel). Within the first picosecond, the signal from Fe 3d spins was significantly reduced while the one from the Gd 4f spins increased dramatically, which was interpreted as non-local angular momentum transfer from the Fe-rich to the Gd-rich regions, illustrated in figure 1c. Although magnetic switching was not directly observed, this process is probably relevant in the understanding of ultrafast magnetization reversal.

Beyond ferro- and ferrimagnetic metallic compounds presented above, time-resolved x-ray scattering techniques are also powerful in clarifying the ultrafast dynamics in materials with antiferromagnetic ordering. Resonant x-ray scattering[2,65,66] (RXS), recently extended to the time domain, has emerged as powerful tool to follow these dynamics, with element specificity and with sensitivity on nanometer length scales. The first reported femtosecond RXS experiment, performed by *Holldack et al.* in the magnetic semiconductor EuTe at the Eu M-edge, demonstrated that the optical excitation of 4f→5d transition reduces antiferromagnetic order on the europium sites with a time constant of <700 fs[67]. It was speculated that exchange interactions were modified as the result of optically induced lattice deformations that happen on the time scale of acoustic motion.

More recently, *Rettig et al.* employed time-resolved RXS to probe magnetization dynamics in antiferromagnetic holmium[68]. On each Ho atom, the total magnetic moment is carried mostly by the localized 4f electrons and only partially by the itinerant, valence band forming 5d electrons. Time-resolved resonant magnetic scattering at different atomic transitions was used to reconstruct separately the dynamics of the localized 4f and 5d spins when near-infrared femtosecond pulses selectively excited the 5d electrons. The experiment showed that the 4f-5d exchange coupling is so strong that the spins on these different electrons were quenched on the same time scale.

Lastly, magnetic scattering allows for imaging magnetization dynamics using x-ray holography[69]. In a first experiment, *Wang et al.* demonstrated the possibility of collecting high-quality magnetic holograms using femtosecond soft x-ray pulses[70]. More recently, *von Korff Schmising et al.* used the same technique to directly image ultrafast demagnetization dynamics at domain wall boundaries in a Co/Pd compound[71] confirming the previous observations of *Pfau et al.*[61]. More recently, *Seaberg et al.* combined x-ray photon correlation spectroscopy (XPCS) with coherent resonant magnetic x-ray scattering to study spontaneous fluctuations of magnetic Skyrmions on nanosecond time scales.[72] These results illustrate that x-ray photon correlation spectroscopy can now study excitations in the μeV energy range, possibly making it complementary to inelastic x-ray and neutron scattering.

The area of femtosecond magnetism has hence made extensive use of ultrafast x-ray sources. In addition, these experiments have been effective in developing a whole set of new techniques, which will impact this and other areas of non-equilibrium dynamics as FEL sources improve in quality and grow in number.

*Charge, orbital, spin and lattice dynamics in complex oxides*

Many transition-metal oxides with fractionally filled d-shells exhibit interesting collective phenomena that descend from strong electronic correlations. Especially, new phases emerge that are not captured by the familiar concepts of band theory and classical magnetism. These phases are also delicate, in that they can be easily switched by external stimulation, for example with static magnetic or electric fields and hydrostatic pressure. Excitation with light can also tip the balance between stable phases, sometimes switching the electronic properties on ultrafast time scales. As discussed above in the magnetism section, femtosecond resonant soft x-ray scattering (RSXS) is a natural tool to study the evolution of these electronic or magnetic degrees of freedom. In combination with time-resolved hard x-ray scattering, which tracks atomic structural rearrangements, or terahertz spectroscopy that measures the optical conductivity, time resolved RSXS determines directly how the electronic ordering is affected and through which stages the phase transition occurs.

For example, in the photo-irradiated magnetite ($Fe_3O_4$)[73,74] time-resolved Fe $L_3$-edge RSXS was used to track the evolution of the electronic order after excitation. The dynamical changes observed in these measurements were interpreted in terms of the breaking of iron trimerons followed by mobile charge creation[9].

Resonant diffraction can also be applied at the metal K-edges to measure charge and orbital order on the metal sites in complex oxide compounds[75,76]. When combined with non-resonant scattering, this technique can be used to measure the dynamical interplay between charges and the lattice, as demonstrated by *Beaud et al.* in the case of the optically induced insulator-metal transition in the manganite $Pr_{0.5}Ca_{0.5}MnO_3$[77]. In this material, the equilibrium charge order is connected to long-range orbital order and to a Jahn-Teller distortion of the crystal lattice[78]. Time-resolved x-ray scattering techniques were used to track the time evolution of these orders. Both near-infrared optical excitation and mid-infrared excitation of the Mn-O stretching vibrations[79] were shown to melt the charge order, followed by relaxation of a Jahn-Teller distortion and of the orbital order.

Time-resolved RSXS experiments were performed in the single-layer manganite $La_{0.5}Sr_{1.5}MnO_4$, in which charges, orbitals, and spin form the so-called CE-type ordering pattern[80]. While charge carrier excitation in the near-infrared perturbed short-range spin ordering very effectively[81], the long-range Jahn-Teller distortions and resulting orbital order was only weakly affected[82].

Mid-infrared resonant driving of a Mn-O lattice mode was also shown to perturb spin and charge/orbital order in $La_{0.5}Sr_{1.5}MnO_4$[83]. This result was interpreted in the context of nonlinear lattice dynamics[84-86], where the crystal lattice is displaced along the coordinates of an anharmonically coupled Jahn-Teller mode to exert a force on the spin and orbital order. Similar experiments were performed in the layered nickelate $La_{1.75}Sr_{0.25}NiO_4$[87,88], where charges and spins order in stripes within the Ni-O planes. Time-resolved RSXS studies at the Ni L-edge revealed that charge and spin order melt also in this material when excited by near-infrared electronic[89,90], as well as mid-infrared lattice excitation[91].

In multiferroic materials (anti-)ferromagnetic order and ferroelectricity coexist and interact by magnetoelectric coupling[92-94]. Coherent spin control by intense THz electric field pulses resonantly driving an electromagnon was demonstrated in multiferroic $TbMnO_3$ [95]. A time-resolved RSXS experiment highlighted such multiferroic control on the sub-picosecond level, providing another new perspective for high-speed optical data storage devices.

*Johnson et al.* showed that short-pulse optical excitation could also be used to tip the balance between different magnetically ordered states[96]. CuO exhibits a lattice-commensurate (CM) collinear antiferromagnetic order in the ground state, but non-collinear incommensurate (ICM) antiferromagnetism in a multiferroic state at intermediate temperatures around 220 K[97]. While the two phases typically coexist in different domains, near-infrared excitation induced a partial change of the average magnetic order in favor of the ICM state. Strikingly, this phase transition sets in after an intensity dependent delay. It was speculated that the magnetic phase transition might be mediated by acoustic-branch magnetic excitations in analogy to the way structural phase transition are mediated by phonons. As the energy barrier between the ICM and CM phase decreases with increasing excitation energy the delay approaches the lower limit of 400 fs, equivalent to one quarter of the oscillation period of a spin wave in this material.

Recently, *Langner et al.* used time-resolved RSXS to study the spin-scattering dynamics in the skyrmion and conical phases of $Cu_2OSeO_3$ upon excitation with near-infrared and ultraviolet pulses.[98] This material shows a complex phase diagram, with a rich number of competing phases that exhibit different magnetic structure,[99] where the application of an external magnetic field to the conical phase creates a skyrmion phase. The skyrmions are topologically protected spin configurations that have recently attracted attention for their robustness to external perturbations and for their potential in data storage applications[100]. Interestingly, the work showed that the skyrmion phase is more robust to optical excitation than the conical phase, possibly due to different spin scattering processes involved in the two phases.

*Heterostructures*

Complex oxide heterostructures have attracted significant interest over the past years, since interfacial coupling allows one to manipulate the static electronic and magnetic material properties and to create new functionalities at equilibrium[101]. A striking extension of these physics to dynamical settings started when mid-infrared light fields, made resonant with specific phonon modes of the substrate, were used to trigger interfacial distortions and to modify the electronic properties of functional films dynamically. This was vividly demonstrated by *Caviglia et al.*, who showed that large-amplitude excitation of the Al-O stretch mode in a $LaAlO_3$ substrate induces an ultrafast insulator-metal transition in a $NdNiO_3$ thin film grown on top[102]. Detailed insight into the spatiotemporal evolution of the different degrees of freedom of this "ultrafast strain engineering" phenomenon was obtained in a set of resonant and off-resonant x-ray diffraction experiments by *Först et al.* where the concomitant dynamics of antiferromagnetic order[103], charge disproportionation and lattice dynamics[104] were investigated. Figure 2a shows the dynamic change of the (¼ ¼ ¼) diffraction peak measured in a time-resolved RSXS experiment at the 852-eV Ni $L_3$-edge. The observed peak reduction and concomitant broadening indicate heterogeneous melting of antiferromagnetic order. Figure 2b plots the time evolution of the intensity of the (2½ 2½ 2½) peak measured on and off resonance with the 8.34-keV Ni K-edge. The on-resonance intensity comprises a charge order contribution (illustrated by the gray shaded region), which disappears on a time scale shorter than that of the off-resonant intensity, which is sensitive to purely structural dynamics. Combining these different measurements leads to the data shown in Figure 2c. The lattice, magnetic and insulator-metal dynamics are illustrated through different types of order-disorder fronts that were found to propagate from the interface into the functional film at different speeds, with charge order melting presumably being the driving force and advancing supersonically ahead of demagnetization and structural relaxation. A sketch of these propagation fronts in the heterostructure is shown in Figure 2d.

*Charge Density Wave Materials*

Charge density waves (CDW) are periodic modulations of the valence electron density in materials. CDWs emerge in materials with strong electron-phonon coupling, for which the total electron energy is reduced by a periodic modulation of the crystal lattice, resulting in a stable or dynamical pattern for low enough temperatures. As a result, a small energy gap forms at the Fermi energy at the wave vector $q$ of the periodic modulation. Charge density waves exhibit peculiar electrical properties, such as nonlinear currents in response to AC and DC electric fields, which have motivated significant interest for electronic device technology over the past several decades[105,106]. When driven out of equilibrium, the structural and electronic degrees of freedom, which are intertwined at equilibrium, may decouple and respond differently on ultrafast timescales[107-111]. Amongst the most studied cases of photo-induced dynamics in CDW materials we mention here $K_{0.3}MoO_3$ (blue bronze), which exhibits a one-dimensional CDW along otherwise metallic chains of corner-sharing $MoO_6$ octahedra. Above a certain light intensity threshold, the excitation of blue bronze with ultrashort infrared and visible pulses melts the charge density wave order on femtosecond timescales [112-115]. Ultrafast x-ray studies reported by *Huber et al.* were interpreted by positing a prompt reshaping of the lattice potential as a cause for a structural relaxation within 100 fs[116] that is able to launch coherent oscillations of the CDW amplitude mode. Indications of a more complex interplay between structural and electronic degrees of freedom were also found when the response of the CDW order was compared for optical excitation of either the electronic subsystem or the crystal lattice with mid-infrared pulses[117]. In both cases, the onset of melting was found above the same threshold magnitude of the coherently driven amplitude mode oscillations, highlighting the existence of a universal stability limit for charge density waves, reminiscent of the Lindemann criterion for the melting of a crystal lattice. In view of their competition with other types of orders, most prominently Cooper pairing in high-$T_C$ superconductors, and their role in the formation of emergent functionalities, the study of charge density waves remains a key aspect in ultrafast research.

### *Light-induced superconductivity in the cuprates*

Hole doped cuprates of the type $YBa_2Cu_3O_{6+x}$ are a family of high-$T_C$ superconductors with crystal structure as sketched in Fig. 3a. Coherent tunneling of Cooper pairs between adjacent $CuO_2$ bilayers along the crystal $c$ axis makes the coherent transport below the critical temperature three-dimensional. Superconductivity is strongly enhanced by a reduction of the distance $d$ between apical oxygen and planar copper atoms, as achieved at equilibrium by the application of pressure in the range of a few kbar[118-121]. This relation opens up interesting opportunities for the control of superconductivity with light. Mid-infrared pulses were used to resonantly excite large amplitude oscillations in this distance $d$, which induced picosecond-lived signatures of coherent transport above the critical temperature and even up to room temperature[12,122]. The underlying dynamics of the crystal lattice were only recently clarified in time-resolved x-ray diffraction experiments by *Mankowsky et al.* at the LCLS free electron laser[123]. Light-induced changes in the intensity of selected Bragg reflections sensitive to the motion of Cu and O ions along the $c$ axis, were measured to identify transient atomic rearrangements locked to the appearance and decay of the transient superconducting state (Figure 3b). In the framework of nonlinear lattice dynamics[84], anharmonic coupling of the resonantly driven Cu-O stretch mode to Raman-active lattice modes was expected to displace the crystal lattice quasi-statically along the coordinates of the latter.

Figure 3c depicts the key element of the photo-induced motions in $YBa_2Cu_3O_{6.5}$ at 100 K, a transient reduction of the important apical-oxygen planar-copper distance $d$. In analogy with static pressure-induced effects, this motion might facilitate the light-induced coherent Cooper pair transport along the crystal $c$ axis above the equilibrium critical temperature ($T_C$ = 50 K), as identified by density functional theory calculations of the electronic structure in the transient state[123]. Among other effects, they predicted the transfer of electrons from the $CuO_2$ planes to the Cu-O chains, similar to hole doping. This interpretation was supported by femtosecond resonant soft x-ray absorption experiments[124].

In high-$T_C$ cuprates with doping levels close to 12.5%, superconductivity competes with the ordering of charges to reduce the critical temperature. The best-known examples are two-dimensional charge density waves in bi-layer $YBa_2Cu_3O_{6+x}$ for $x \approx 0.6$[125,126] and charge stripes in the single-layer $La_{2-x}Ba_xCuO_4$ compounds for $x = 0.125$[127,128], with the latter illustrated in Figure 4a. A key question in understanding light-induced superconductivity is how these competing orders evolve when the transient superconducting states are formed from the charge ordered state.

*Först et al.* used time-resolved RSXS at the oxygen K-edge to find prompt and complete melting of the stripe order in the frustrated superconductor $La_{1.875}Ba_{0.125}CuO_4$ that was illuminated by intense mid-infrared pulses resonant with the in-plane Cu-O stretch mode[129]. Importantly, the same excitation in the closely related compound $La_{1.675}Eu_{0.2}Sr_{0.125}CuO_4$ induces transient superconductivity, as probed by time-resolved THz spectroscopy[11]. The combination of the two experiments, as shown in Figure 4b and 4c, strongly suggests that melting of the competing stripe order is prerequisite for the formation of the transient coherent state. A similar result was found in $La_{1.885}Ba_{0.115}CuO_4$, where for certain temperatures superconductivity and stripe order co-exist at equilibrium[128]. The photo-induced destruction of charge order, measured by time-resolved RSXS, appears to be concomitant with the dynamical enhancement of the superconducting order observed in time-domain THz spectroscopy[130].

This dynamical interplay of competing orders was also observed in $YBa_2Cu_3O_{6.6}$ above the equilibrium critical temperature. The resonant optical excitation of the apical oxygen vibrational mode, which induces out-of-plane interlayer coherence as discussed above, partially melts the in-plane charge-density wave order as identified in a time-resolved RSXS experiment at the 932 eV Cu $L_3$-edge[131].

For the first time, x-ray scattering measurements succeeded to capture the crystallographic and the electronic properties of a transient room temperature superconductor. These findings may illustrate new pathways towards the design of novel materials exhibiting equilibrium room temperature superconductivity.

*Electron phonon coupling in high-$T_C$ superconductors*

As demonstrated in many cases above, combining direct measurements of the lattice structure with those of the electronic degree of freedom provides new important insights into the emergence of exotic states of matter. Information from complementary experiments such as time-resolved x-ray diffraction and time-resolved angular resolved photoemission spectroscopy (trARPES) can clarify the strength and origin of electron-phonon coupling in complex materials. *Yang et al.* observed global oscillations of the Fermi level in $BaFe_2As_2$, a parent compound of FeAs based high-$T_C$ superconductors, at the frequency of the $A_{1g}$ phonon mode, which themselves suggested the presence of a strong electron phonon coupling[132]. These results were then complemented by ultrafast x-ray diffraction experiments[133,134] that quantified how the coherent excitation of the $A_{1g}$ mode modulates the Fe-As-Fe bond angle. In combination, these measurements lead to an estimate of the electron-phonon deformation potential and coupling constant for the $A_{1g}$ mode, which was in good agreement with density functional theory calculations.

More recently, it has been proposed that electronic correlations strengthen electron-phonon coupling in iron selenide and iron pnictides superconductors and may play a role in the emergence of superconductivity in these materials[135]. In a pioneering experiment, *Gerber et al.* were able to quantify the electron-phonon coupling in FeSe superconductors[136]. Photo-excitation of FeSe with 1.5eV femtosecond pulses triggered a coherent oscillation of the $A_{1g}$ mode. Time resolved hard x-ray diffraction tracked the motion of the selenium atom $\delta Z_{Se}$ shown in figure 5a, while high-resolution time-resolved ARPES tracked the shift of the $d_{xy/yz}$ and $d_{z2}$ orbital bands shown in figures 5b and 5c, respectively. These measurements yielded orbital-resolved values for the electron-phonon deformation potential that could be compared directly to theory predictions. The retrieved experimental values could only be captured when DFT was combined with dynamical mean field theory to include electron-electron correlation effects, proving their importance in determining the electron-phonon coupling in FeSe and related materials.

*Probing Elementary Excitations in the Time Domain*

Probing low-energy excitations in solids and their dispersion reveals important information on the fundamental interactions at play. Resonant inelastic x-ray scattering (RIXS) is a photon-in photon-out technique that allows for the study of elementary excitations in solids at finite momenta with orbital and element selectivity[137]. Several elementary excitations can be probed using RIXS such as charge transfer excitations and d-d transistions[138,139], magnons in 2D or 3D[140,141] and phonons[142,143]. Time-resolved RIXS would enable reconstructing the time evolution of such excitations, for example when the material is driven out of equilibrium by a short laser pulse. In a pioneering experiment *Dean et al.* employed time-resolved RIXS to study how magnetic correlation evolve upon photo-doping the $Sr_2IrO_4$ Mott insulator[144]. The 3D magnetic order, measured by time-resolved RSXS, was completely quenched within the first 2 ps and recovered with a fluence dependent time constant that varies between 100 ps and 1 ns. Time-resolved RIXS then examined the evolution of magnetic excitations and revealed that 2D in-plane magnetic correlations recover on a much faster time scale, which is similar to that of charge recombination. It was speculated that such fast recovery time might be due to the fundamental link that exists in strongly correlated materials between the in-plane electron hopping parameter and the in-plane magnetic exchange. The slow recovery of the long-range magnetic order was instead related to the weak inter-plane exchange coupling and energy dissipation into other degrees of freedom.

Short x-ray pulses also allowed for inelastic measurements of non-equilibrium lattice dynamics. *Trigo et al.* demonstrated a highly interesting and new experimental technique to probe phonon dispersion curves in solids by measuring time resolved x-ray scattering[145]. In this experiment, a germanium single crystal was excited with near-infrared laser pulses to produce correlated phonon pairs at opposite momenta that modulate the x-ray diffuse scattering intensity around Bragg peaks at twice the phonon frequency. In contrast to typical x-ray measurements that analyze the incoherent thermal diffuse scattering and require an *ab-initio* model of the interatomic forces, this method allows to extract the phonon dispersion curves directly from Fourier transformations of the modulated diffuse scattering intensity. Figures 6b and 6c show the obtained dispersion relations for two transverse acoustic modes in germanium along the directions sketched in Figure 6a. The agreement with the calculated dispersions (white lines) is good, especially considering that there are no adjustable parameters. *Jiang et al.* employed this technique to investigate the origin of incipient ferroelectricity in PbTe. They found that the ferroelectric instability is due to the existence of strong electron-phonon interactions rather than phonon-phonon anharmonicities[146].

Probing excitations in the time domain also comes with significant advantages over equilibrium measurements such as inelastic neutron and x-ray scattering. For example, Fourier transform inelastic x-ray scattering in the time-domain allows easier access to the lower frequency part of the dispersion relations of an excitation. Furthermore, while traditional inelastic measurements only reveal harmonic properties of phonons in a momentum-resolved manner, time-resolved x-ray scattering can reveal the presence of anharmonic couplings between different phonon modes. In a recent experiment,

*Teitelbaum et al.* succeeded to directly identify the individual decay channel of the $A_{1g}$ phonon mode in bismuth[147]. In this experiment, they also obtained, for the first time, a quantitative measurement of the anharmonic force constants between the $A_{1g}$ mode and the anharmonically-coupled longitudinal acoustic modes.

*Perspectives and Conclusions*

All the experiments discussed above show how the advent of short x-ray pulses has enabled a far deeper understanding of non-equilibrium phenomena in complex solids. In most cases, free-electron laser operation is based on self-amplified spontaneous emission (SASE), producing pulses with a large bandwidth and strong shot-to-shot fluctuations in most of their key parameters, such as intensity, duration, and spectrum[40]. While one can often account for these fluctuations using single-shot diagnostics,[148-150] the large bandwidth of the x-ray pulses severely limits the energy resolution achieved during experiments. This problem is typically mitigated by the use of monochromators, with the caveat that they cause a severe reduction (up to two orders of magnitude) in the available x-ray intensity.

To overcome this limitation, a key aspect in the design of advanced FEL sources is the adoption of seeding schemes with the ultimate aim of reaching Fourier transform-limited x-ray pulses.[151-153] Compared to SASE operation, seeded FEL operation produces pulses with narrower bandwidth and significantly improved energy and intensity stability. Another important benefit of newly developed FEL sources is the increase in pulse repetition rate that will allow collecting data at reduced experimental time. In combination, we expect that seeded FEL sources with high repetition rates will foster a number of flux hungry techniques, such as time-resolved resonant x-ray diffraction or resonant inelastic x-ray scattering, with even higher time and spectral resolution.

Finally, these new sources promise the complete control of the x-ray pulse parameters, aiming to reach sub-fs duration with arbitrary pulse shape. With more precise synchronization between the x-ray pulses and the optical excitation fields, for example in the case of THz and mid-infrared pulses,[154] it is already becoming possible to make the x-ray pulses stable with respect to the absolute phase of the pump pulses. We envision that this advancement will open up entirely new areas of research investigating coherence effects. For example, it will be possible to follow changes in the material structure and electronic properties as they are induced by the excitation electric field, promising further insight into the origin of nonlinear couplings between different excitations in condensed matter.

**Acknowledgements**

We acknowledge funding from the European Research Council under the European Union's Seventh Framework Programme (FP7/2007-2013)/ERC Grant Agreement no. 319286 (QMAC). We acknowledge support from the Deutsche Forschungsgemeinschaft via the excellence cluster "The Hamburg Centre for Ultrafast Imaging - Structure, Dynamics and Control of Matter at the Atomic Scale" and the priority program SFB925. M. B. acknowledges financial support from the Swiss National Science Foundation through an Early Postdoc Mobility Grant (P2BSP2_165352).

## Box 1 – Time Resolved X-ray Techniques

Pump-probe x-ray techniques constitute a valuable tool to reconstruct material dynamics by capturing directly the transient light-induced changes in the microscopic degrees of freedom. A typical x-ray time-resolved experiment is illustrated in the figure below. The sample under study is excited with a strong laser pulse, identified as pump, that triggers dynamics in the material. A time-delayed x-ray pulse probes the pump-induced changes by interaction with the material and subsequent collection of the scattered (or transmitted) beam on a detector. Depending on the choice of the photon energy in the hard and soft x-ray regimes, information about the atomic or electronic structure of the material can be retrieved using techniques such as x-ray diffraction, x-ray absorption spectroscopy and resonant x-ray diffraction.

Discovered by Max von Laue in 1912, x-ray diffraction is arguably one of the most useful tools in material characterization. X-ray waves scattered by the periodically ordered atoms in a crystal interfere constructively or destructively along specific directions. The analysis of the measured interference patterns allows for determining the average position of each atom in the crystal with sub-picometer spatial resolution. For example, by monitoring the position and relative intensities of a purposely chosen set of Bragg peaks as a function of time, one can trace lattice dynamics triggered by the excitation of a coherent phonon[25-27]. With the symmetry of the material determining the measured diffraction pattern, one can also follow through which stages an ultrafast light-induced phase transition occurs[28]. As a last example, in analogy to the case of thermal diffuse scattering, the time-dependence of the scattered intensity in-between Bragg peaks reveals information about the dispersion of phonons without the need of *ab-initio* modeling of force constants[145].

X-ray absorption spectra contain fingerprints of materials' electronic and magnetic structures. As the x-ray energy is tuned to resonance with an atomic transition, the absorption increases dramatically and electronic and magnetic states can be reconstructed in an element-specific manner. By analyzing how the absorption spectrum of a substance changes upon photo-excitation, it is possible, for example, to gain insight into transient changes in the oxidation states and bond lengths of a compound[124,155]. Also, changes in the magnetic moment of a specific atom can be disclosed by measuring x-ray magnetic circular dichroism under resonance condition[156].

Such absorption spectroscopy captures sample properties without the possibility of spatial reconstruction, making it difficult to observe complex long-range ordering of charges, spins, and orbitals. Resonant x-ray diffraction combines the contrast mechanisms given by absorption spectroscopy with the spatial resolution given by diffraction. By performing diffraction experiments with incoming photons tuned in resonance with appropriate atomic transitions it is possible to study directly phenomena such as charge stripe order in cuprates[128,129] or orbital and spin order in manganites[75,77]. In recent years, also this technique has been brought into the time domain and has become a standard experiment for materials research at x-ray FELs.

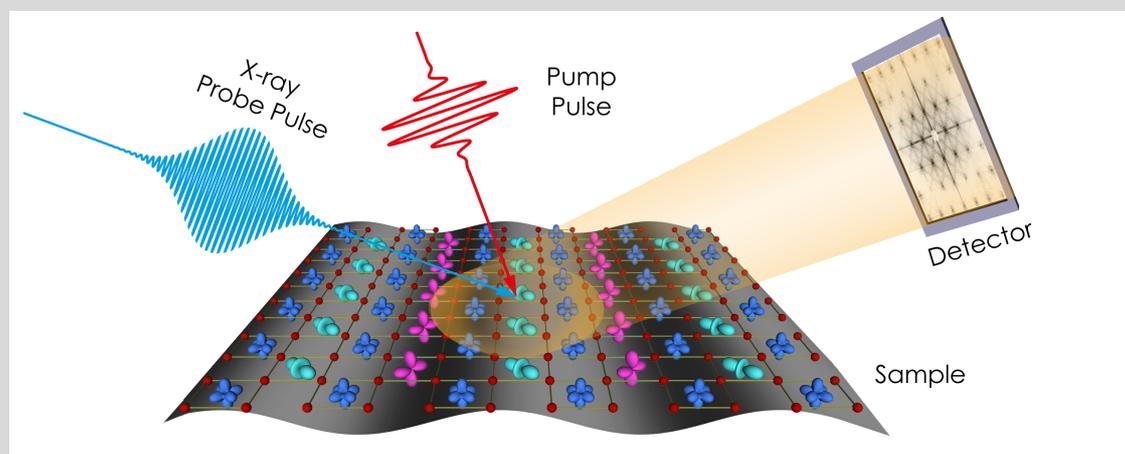

# Figures

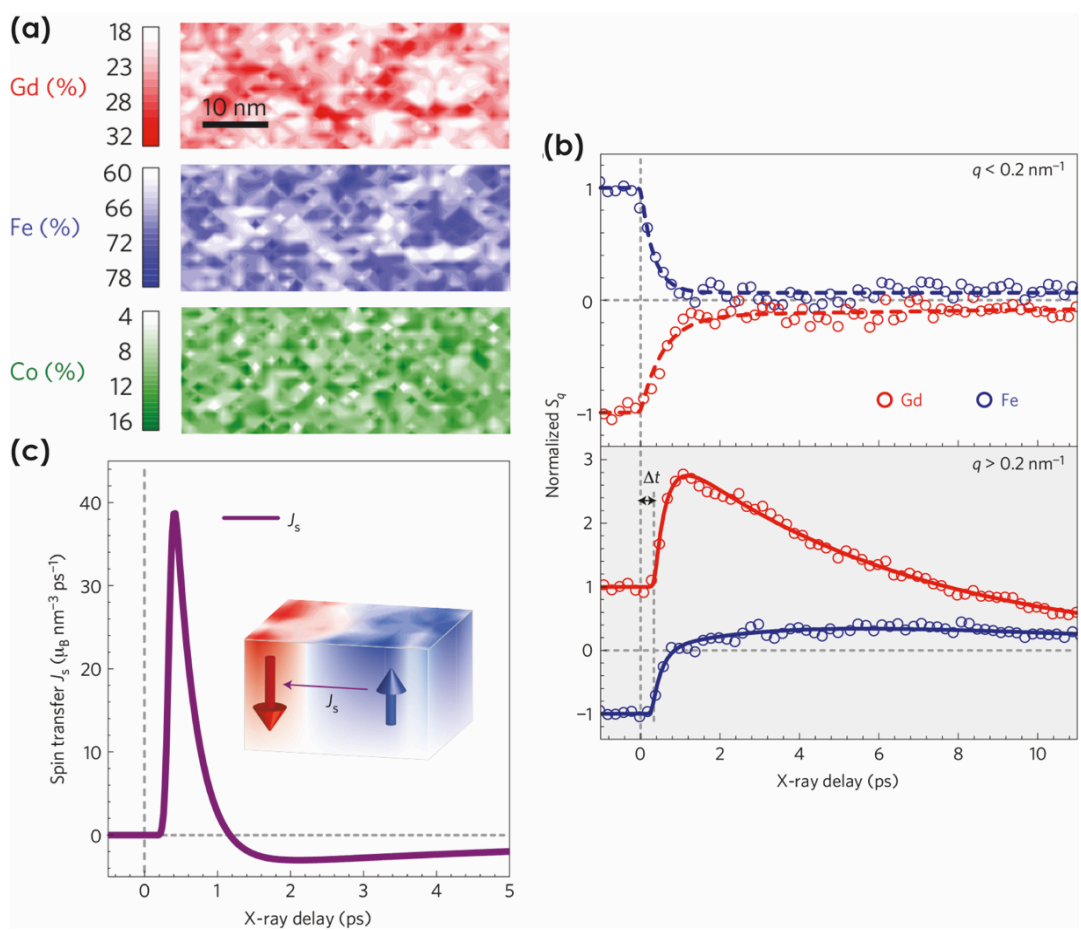

Fig. 1| **Ultrafast angular momentum transfer in a ferromagnetic film.** Non-local, ultrafast transfer of angular momentum in GdFeCo was revealed by time resolved magnetic small angle X-ray scattering[5]. **a|** Local chemical nanoscale variations for Gd, Fe and Co in $Gd_{24}Fe_{66.5}Co_{9.5}$ as measured with energy dispersive X-ray spectroscopy. Darker coloured areas indicate elemental enrichment, whereas white areas indicate below average concentrations. **b|** Temporal evolution of the magnetic diffraction, $S_q$, for Gd 4$f$ (red) and Fe 3$d$ (blue) spins. The time delay ($\Delta t$) is defined as the time interval between the arrival of the pump pulse and the x-ray probe pulse. The measurement of $S_q$ at low-scattering momenta ($q$) (top part) probes the average sample magnetization, whereas the high-$q$ $S_q$ (bottom part) measures the evolution of the magnetization at the nanoscale, showing the transfer of angular momentum to the Gd-rich regions. **c|** Time evolution of the angular momentum transfer to the Gd-rich regions as extracted from the analysis of the high-$q$ $S_q$ scattering data. The ultrafast transfer of angular momentum takes 1 ps and is followed by a slow recovery through spin dissipation within the Gd regions. Adapted with permission from *Graves et al.* [5], Nature Publishing Group.

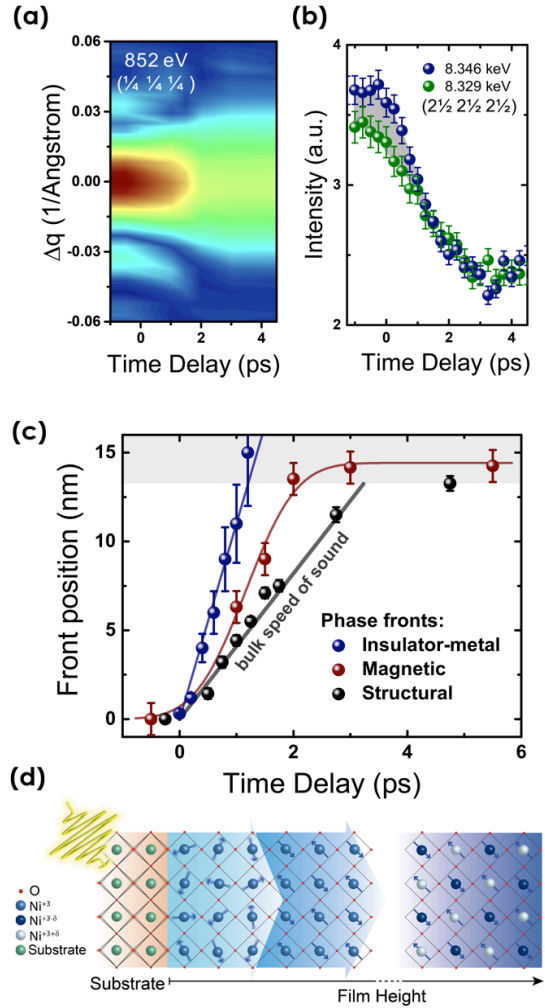

Fig. 2| **Ultrafast interfacial strain engineering.** In a LaAlO$_3$/NdNiO$_3$ heterostructure, the dynamics of multiple magnetic, electronic and structural degrees of freedom in the NdNiO$_3$ thin film can be observed when an insulator–metal transition is driven by resonant excitation of a high-frequency vibrational mode in the underlying LaAlO$_3$ substrate. **a|** Transient momentum-resolved intensity of the (¼ ¼ ¼) diffraction peak measured at the 852 eV Ni L$_3$-edge, which is sensitive to antiferromagnetic ordering. **b|** Transient peak intensity of the NdNiO$_3$ (2½ 2½ 2½) reflection measured at X-ray energies resonant (8.346 keV, blue) and off-resonant (8.329 keV, red) with the Ni K-edge. The measured resonant diffraction intensity includes a charge order contribution (grey shaded region), which disappears on a time scale shorter than that of the off-resonant intensity and is sensitive only to structural dynamics. **c|** Spatiotemporal evolution of the NdNiO$_3$ lattice, magnetic and insulator–metal dynamics along the thin film out-of-plane direction, extracted from the time-resolved diffraction experiments. Individual phase fronts of insulator–metal, antiferromagnetic–paramagnetic and structural transitions propagate at different speeds from the LaAlO$_3$/NdNiO$_3$ heterointerface into the nickelate film. **d|** Illustration of these dynamics induced by the substrate phonon excitation. Panel **(a)** adapted with permission from *Först et al.,*[103] Nature Publishing Group. Panels **(b)**, **(c)** adapted with permission from *Först et al.*[104], American Physical Society.

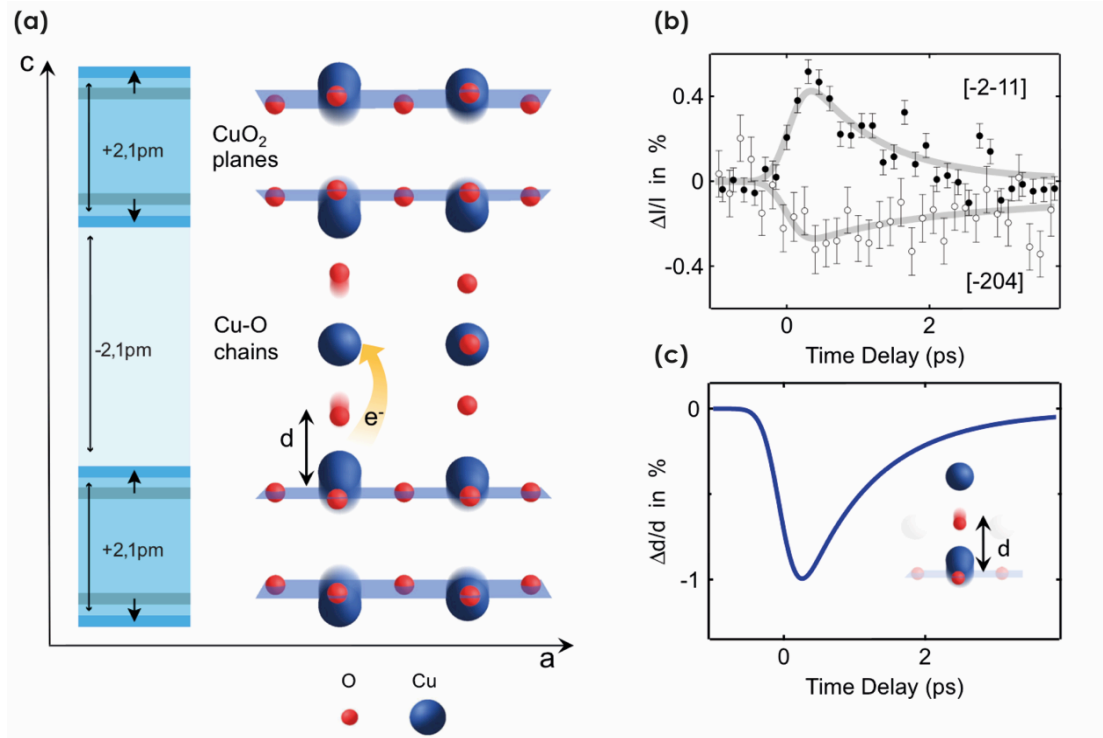

Fig. 3| **Nonlinear phononics in a bilayer cuprate.** Signatures of nonlinear phonon coupling in the high-temperature superconductor $YBa_2Cu_3O_{6.5}$ can be identified upon driving the $B_{1u}$ phonon mode to large amplitudes with 15 µm ultrashort laser pulses. **a|** Schematic illustration of the crystal structure of bilayer $YBa_2Cu_3O_{6.5}$ (right). $CuO_2$ bilayers (grey) are oriented perpendicular to the *c* axis and alternate with thicker layers containing Ba, Cu and O. Y and Ba atoms are not shown for clarity. *d* is the distance between an apical O atom and a Cu atom in the plane. Nonlinear coupling of the driven $B_{1u}$ mode to $A_g$ modes induces changes in the intra-bilayer (+2.2 pm) and inter-bilayer (–2.2 pm) distances (left). Light and dark grey represent the $CuO_2$ planes in the equilibrium and driven positions, respectively. **b|** Transient intensity (*I*) of two exemplary Bragg peaks. The solid curve is obtained from the ab initio calculated structure by considering nonlinear phonon coupling of the driven $B_{1u}$ mode to the $A_g$ modes. **c|** Time-resolved change in the O–Cu distance *d* obtained from the calculated structure. Adapted with permission from *Mankowsky et al.* [123], Nature Publishing Group.

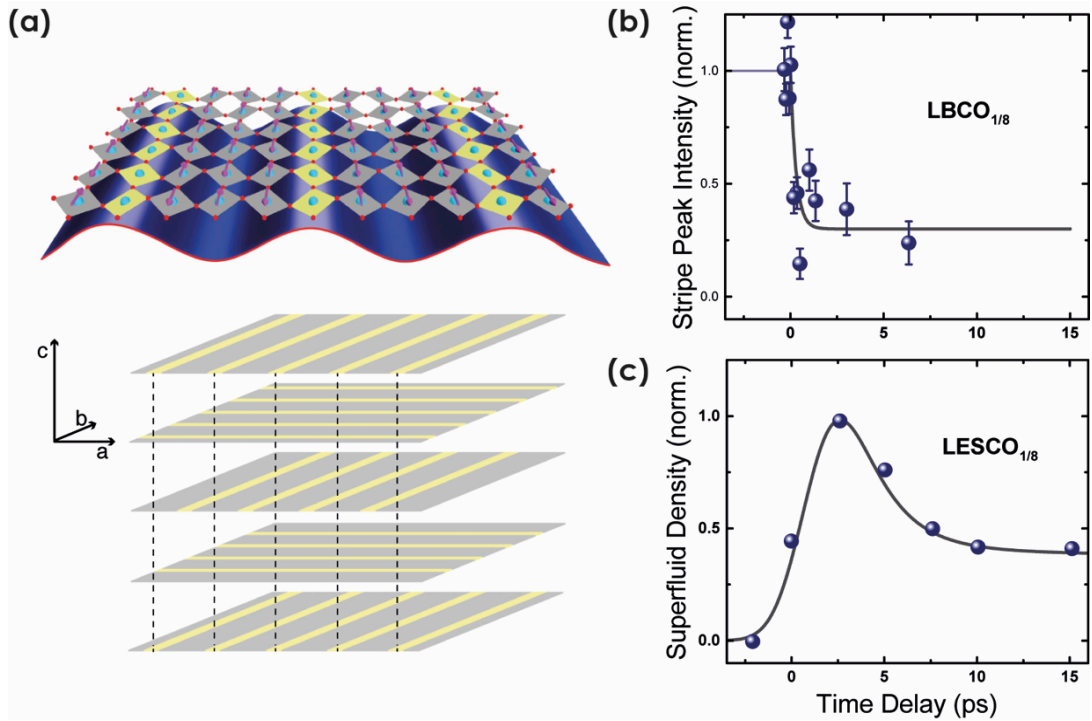

Fig. 4| **Ultrafast stripe order melting in a single-layer cuprate.** Stripe order melting in the single-layered high-temperature superconductor $La_{1.875}Ba_{0.125}CuO_4$ ($LBCO_{1/8}$) can be triggered by driving an in-plane Cu–O stretching mode to large amplitudes with 14.5 µm ultrashort laser pulses. **a|** Illustration of the in-plane stripe order in single-layered cuprates and the 1D modulation of charges and the spins responsible for suppressing superconductivity at equilibrium (top), emphasizing the buckling in the Cu–O planes (Cu atoms are blue, O atoms are red and the spins are represented by arrows). The stripes are periodically stacked along the $c$ axis with a 90° rotation between adjacent layers (bottom). **b|** Transient intensity of the stripe order diffraction peak at the (0.24 0 0.5) wave vector, measured in $LBCO_{1/8}$ at resonance with the O K-edge at 528 eV. **c|** Evolution of the normalized superfluid density in $La_{1.675}Eu_{0.2}Sr_{0.125}CuO_4$, ($LESCO_{1/8}$), a compound closely related to $LBCO_{1/8}$, in which excitation of the same Cu–O stretching mode induces a transient superconducting state. Panels **(a)**, **(b)** adapted with permission from *Först et al.*[129], American Physical Society. Panel **(c)** adapted with permission from *Fausti et al.,*[11] AAAS

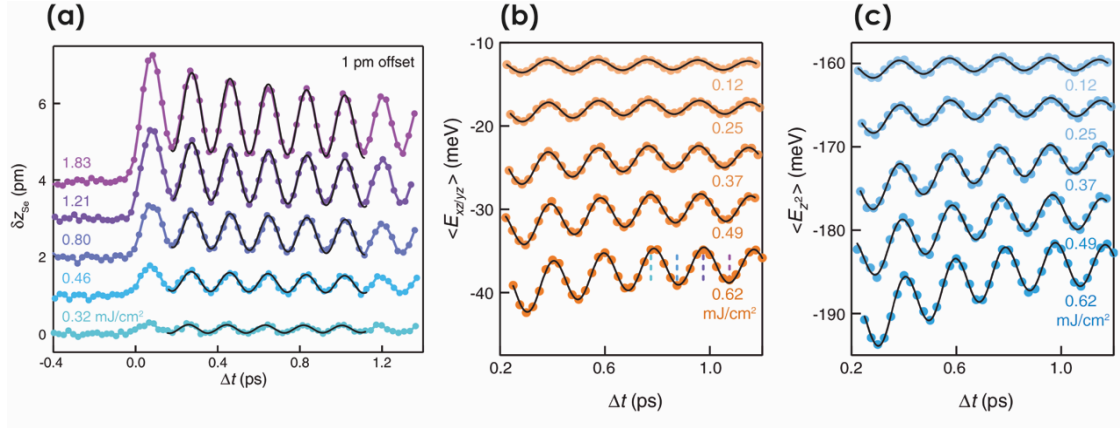

Fig. 5| **Electron–phonon deformation potential in an iron-based superconductor.** The orbital-resolved electron–phonon deformation potential in an FeSe superconducting thin film was determined using time resolved X-ray diffraction and time-resolved angular-resolved photoemission spectroscopy (tr-ARPES). **a|** Displacement of the Se atom, $\delta Z_{Se}$, extracted from the evolution of the intensity of the (004) Bragg peak. **b, c|** Momentum-averaged energy shifts $<E>$ of the $d_{xz/yz}$ and $d_{z^2}$ bands extracted from tr-ARPES measurements. All measurements are shown for different photoexcitation levels, ranging from 0.12 to 1.83 mJ cm$^{-2}$. Adapted with permission from *Gerber et al.*[136], AAAS.

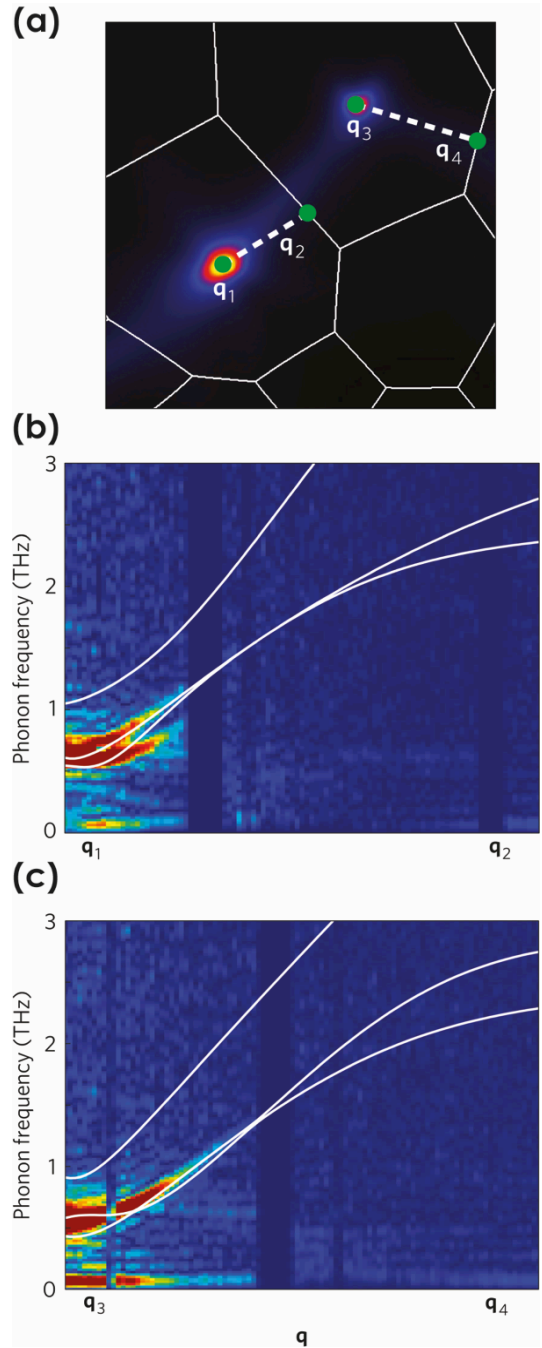

Fig. 6| **Phonon dispersion relation in germanium.** The dispersion relations of the two transverse acoustic branches of Ge were obtained using Fourier-transform inelastic X-ray scattering. The calculated equilibrium diffraction pattern is shown in panel **a**; the dashed lines indicate the directions along which the dispersion relations are shown in panel **b** as a function of the reciprocal space vector $q$, where $q_1 = (-0.1, 0, -0.07)$ and $q_2 = (-0.33, -0.75, 0.37)$, and panel **c**, where $q_3 = (0.13, -0.04, 0.05)$ and $q_4 = (-0.09, -0.98, -0.08)$ (reciprocal lattice units). The solid white lines in panels **b** and **c** represent the calculated acoustic dispersion curves. Adapted with permission from *Trigo et al.*[145], Nature Publishing Group.